\documentclass[useAMS,usenatbib]{mn2e}
\usepackage{graphicx,txfonts,amssymb,amsfonts,natbib,times,hyperref}

\begin{document}

\title[Matter power spectra]
  {Matter power spectra in dynamical Dark Energy cosmologies}

\author[C.\,Fedeli et al.]{C.\,
      Fedeli$^{1}$, K. Dolag$^{2}$, and L. Moscardini$^{3,4,5}$\\
      $^1$ Department of Astronomy, University of Florida, 312 Bryant Space Science Center, Gainesville, FL 32611 (cosimo.fedeli@astro.ufl.edu)\\
      $^2$ Max-Planck-Institut f\"ur Astrophysik, Karl-Schwarzschild Stra\ss e 1, D-85748 Garching, Germany\\
      $^3$ Dipartimento di Astronomia, Universit\`a di Bologna, Via Ranzani 1, I-40127 Bologna, Italy\\
      $^4$ INFN, Sezione di Bologna, Viale Berti Pichat 6/2, I-40127 Bologna, Italy\\
      $^5$ INAF - Osservatorio Astronomico di Bologna, Via Ranzani 1, I-40127 Bologna, Italy}
      
\maketitle

\begin{abstract}
We used a suite of numerical cosmological simulations in order to investigate the effect of gas cooling and star formation on the large scale matter distribution. The simulations follow the formation of cosmic structures in five different Dark Energy models: the fiducial $\Lambda$CDM cosmology and four models where the Dark Energy density is allowed to have a non-trivial redshift evolution. Each simulation includes a variety of gas physics, ranging from radiative cooling to UV heating and supernova feedback (although the AGN feedback is not incorporated). Moreover, for each cosmology we have a control run with dark matter only, in order to allow a direct assessment of the effect of baryonic processes. We found that the power spectra of gas and stars, as well as the total matter power spectrum, are in qualitative agreement with the results of previous works not including the AGN effects in the framework of the fiducial model, although several quantitative differences exist. We used the physically motivated halo model in order to investigate the backreaction of gas and stars on the dark matter distribution, finding that it is very well reproduced by simply increasing the average dark matter halo concentration by $17\%$, irrespective of the mass. This is in agreement with gas cooling dragging dark matter in the very center of halos, as well as adiabatic contraction steepening the relative potential wells. Moving to model universes dominated by dynamical Dark Energy, it turns out that they introduce a specific signature on the power spectra of the various matter components, that is qualitatively independent of the exact cosmology considered. This generic shape is well captured by the halo model if we blindly consider the cosmology dependences of the halo mass function, bias, and concentration. However, the details of the dark matter power spectrum can be precisely captured only at the cost of a few slight modifications to the ingredients entering in the halo model. The backreaction of baryons onto the dark matter distribution works pretty much in the same way as in the reference $\Lambda$CDM model, in the sense that it is very well described by an increment in the average halo concentration.
Nonetheless, this increment is less pronounced than in the fiducial model (only $\sim 10\%$), in agreement with a series of other clues pointing toward the fact that star formation is less efficient when Dark Energy displays a dynamical evolution.\\
\\
{\bf Key words:} cosmology: theory $-$ cosmological parameters $-$ large scale structure of the Universe
\end{abstract}

\section{Introduction}\label{sct:introduction}

Many independent cosmological experiments point nowadays toward a concordance cosmological model in which gas and stars, that is standard baryonic matter, constitute only a few percent of the total energy budget of the Universe. The bulk of this budget is subdivided between a non-baryonic Cold Dark Matter (CDM) component, which decide the dynamics of cosmic structures, and a smooth Dark Energy (DE) component, that affects only the global geometry and expansion rate of the Universe, particularly being responsible for its recent accelerated expansion phase \citep{RI98.1,PE99.1,TE06.1,KO11.1,SU11.1}. The vast majority of cosmological tests are in agreement with a constant DE density (dubbed cosmological constant and indicated with $\Lambda$), hence the standard cosmology is usually labeled as $\Lambda$CDM. Despite its striking observational success, a cosmological constant is largely unappealing from the theoretical point of view, for a variety of reasons that range from quantum gravity considerations to fine tuning and coincidence problems \citep*{CA92.1,PE03.1}.

This fact triggered an enormous amount of work in the past two decades, aimed at finding alternative explanations for the observed geometry and accelerated expansion of the Universe without a cosmological constant. One such alternative consists in considering DE as the energy density of a scalar field (the quintessence) that slowly rolls down a potential well \citep{WE95.1,HE01.1}. This approach would solve many of the theoretical problems of a cosmological constant, however infinitely many such models can be devised, depending on the shape of the scalar field potential and on the nature of its coupling to gravity and matter. As it turns out, having a quintessence model that reproduces the recent accelerated expansion of the Universe within experimental errors is relatively easy, therefore signatures distinctive of quintessence models should be mainly searched for in the formation of cosmological structures.

For the standard $\Lambda$CDM cosmology, the formation of dark matter structures is relatively well understood, thanks to both the linear perturbation theory and numerical cosmological simulations. The latter are of particular importance, in that they allow to follow the complete growth of structures in the deeply non-linear regime, for which semi-analytic methods are unsuited. The reason for this relative success resides in the fact that dark matter is assumed to respond only to gravity, which is easily modeled. Gas physics, on the other hand, is far more complicated, including a plethora of non-gravitational phenomena such as preheating, radiative and metal line cooling, star formation, energy feedback, etc. Many of these processes are active on scales too small to be resolved by current cosmological simulations, thus requiring the adoption of semi-analytic models in order to describe sub-grid physics. Despite being subdominant with respect to dark matter, baryonic matter can have a substantial impact on the distribution of material at small scales (see, e.g., \citealt{PU05.1}), and hence needs to be taken into account when formulating predictions for future high precision cosmological tests.

As mentioned above, given the relevance of the DE problem for modern cosmology, it is important to understand in detail how the entire process of structure formation in quintessence models differs from the case of a mere cosmological constant. This issue has been addressed several times in the past with the use of numerical simulations \citep{DO04.1,BA05.2,GR09.2}, however the focus has always been more on the properties of dark matter structures (abundance, distribution, internal features, etc.) rather than the large scale matter distribution in general. In addition, almost none of these works included standard baryonic matter in the problem. Changes in the properties of dark matter structures are expected to affect baryonic processes, which depend substantially on the environment and on the interplay between the cooling timescale and the dynamical timescale for structure collapse. At the same time, baryons are expected to provide some kind of backreaction on the dark and total matter distributions, which would also depend on the background expansion history of the Universe.

In this paper we thus explored in detail the co-evolution of baryons and dark matter in cosmologies with dynamical DE. We made use of the power spectra of the various matter components, as measured in a suite of hydro-cosmological simulations. The power spectrum includes a wealth of information about the matter density field, and it is the main subject of actual measurements such as cosmological weak lensing. The simulations we used are larger than many, if not all, simulations used before for similar studies, including the mere study of baryons in the concordance $\Lambda$CDM cosmology. As a consequence, they are less affected by finite volume effects, although they conceivably have a lower resolution than a few of the previous works. Large scale structure formation was evolved in five DE cosmologies, namely a fiducial $\Lambda$CDM model, two standard quintessence models and two extended quintessence models. This allowed us to explore differences and regularities amongst models with substantially different redshift evolutions of the DE equation of state parameter. 

In this work we focused exclusively on the simulation snapshots at $z = 0$. As will be seen, our set of results is already significantly dense in that case, thus we believe that a study of the redshift evolution of the matter power spectra would be better suited for a stand-alone effort. The rest of this work is organized as follows. In Section \ref{sct:previous} we summarize the current state of knowledge about how the matter correlation function is affected by gas physics on one hand, and by a background dominated by dynamical DE on the other. This will help to set the stage for what follows, and to better motivate our work. In Section \ref{sct:models} we describe the five DE models (a fiducial $\Lambda$CDM and four dynamical DE) that have been explored in this work, and in Section \ref{sct:simulations} we give some details about the simulations employed. In Section \ref{sct:halo} we describe the halo model, a flexible and physically motivated recipe for computing the dark matter power spectrum that we subsequently used for interpreting our findings. In Section \ref{sct:results} we present our results, while in Section \ref{sct:conclusions} we summarize our conclusions. 

\section{Previous studies}\label{sct:previous}

In this Section we summarized the results of previous studies in the field that we are investigating, in order to better highlight the elements of novelty present in our analysis.

\subsection{Gas physics}

The study of the effect of baryons on the large scale matter distribution in a $\Lambda$CDM cosmology dates back at least to \citet{WH04.1} and \citet{ZH04.1}. In the first paper, the author estimated the changes in the total matter power spectrum due to cooling mechanisms that would lead the gas to concentrate in the very center of dark matter halos, thus dragging along dark matter due to the steepened potential wells. This would lead to an effective increase in the concentration of cosmic structures, translating into an increase in the matter correlation amplitude at $k\gtrsim 5~h$ Mpc$^{-1}$. \citet{WH04.1} used a simple model of adiabatic contraction \citep{BL86.1,KO01.2} in combination with the halo model (see Section \ref{sct:halo} below) in order to replicate this effect, and estimated an increase in total matter power compared to the dark matter only case of $\sim 5-20\%$ at $k \gtrsim 10~h$ Mpc$^{-1}$, depending on the values of his model parameters. \citet{ZH04.1} also employed the halo model, suitably modified in order to include the baryonic contribution to the density distribution of cosmic structures. In this case the authors ignored the baryons that cool down and condense at the center of dark matter halos, while focusing on the hot diffuse baryons, hence finding a decrement of the total matter power spectrum as compared to the dark matter only power spectrum.

Numerical simulations of structure formation have been self-consistently used in order to investigate this issue for the first time by \citet{JI06.1}. Their simulations followed the evolution of $2\times512^3$ gas and dark matter particles within a box of $100~h^{-1}$ Mpc, including radiative cooling, star formation, and supernova feedback. The authors found that the total matter power spectrum was suppressed with respect to the dark matter only case at intermediate scales ($k\sim 1-2~h$ Mpc$^{-1}$) by a few percent, and then very much increased at very small scales ($k\gtrsim 10~h$ Mpc$^{-1}$). This result can be seen as a combination of the effects of cooling baryons described by \citet{WH04.1} and of hot baryons described by \citet{ZH04.1}. Moreover, they were able to study the backreaction of baryons on the dark matter power spectrum, showing that the latter is enhanced at intermediate/small scales with respect to the dark matter only case, e.g.,  by $\sim 10\%$ at $k \sim 10~h$ Mpc$^{-1}$

Subsequently, \citet*{RU08.2} performed an analysis similar to the one of \citet{JI06.1} by adopting $2\times256^3$ dark matter and gas particles in a $60~h^{-1}$ Mpc side box, including gas cooling, star formation, supernova feedback, metal enrichment, and UV heating in their simulations. They found results in qualitative agreement with those of \citet{JI06.1}, although several noticeable quantitative differences exist. Most remarkably, the raise in the clustering strength of the dark matter and total matter at small scales, as compared to the dark matter only spectrum, is much more enhanced, leading to a $\sim 50-70\%$ increase at  $k \sim 10 h$ Mpc$^{-1}$. The authors attribute this difference to the different implementation of gas physics and to finite volume effects, their simulation box being rather small.

By looking for a physical interpretation of the small scale enhancement in the dark matter and total matter power spectra, \citet*{RU08.2} noted that dark matter halos tend to be $\sim 10-20\%$ more concentrated when gas cooling and star formation are included, in a way that is not very much dependent on the mass. Also, the concentration of the total matter distribution inside cosmic structures tends to increase more dramatically, and more so for smaller mass halos, where gas cooling is expected to be more efficient. Hence, motivated by the halo model, they corrected the dark matter only power spectrum by the Fourier transform of a representative dark matter halo density profile having concentration augmented by $\sim 70\%$. They found that this correction reproduces fairly well the total matter power spectrum measured in their simulations. The halo model interpretation of the effect of baryonic physics was also later resumed by \citet*{ZE08.1}, that forecasted the constraints given by tomographic weak lensing on the mass-concentration relation of cosmic structures including gas cooling.

More recently, \citet*{GU10.1} performed yet another similar study by using an even smaller simulation box ($50~h^{-1}$ Mpc on a side), and including radiative and metal line cooling, UV background, and star formation with subsequent supernova feedback and metal enrichment. They found as well a substantial increase in the total matter clustering strength for $k\gtrsim 10~h$ Mpc$^{-1}$, indeed even larger than the one found by \citet*{RU08.2}. In order to mimick this, they not only increased the concentration of dark matter halos, but also included an additional component to the halo model, representing the central baryon condensation. Interestingly enough, both \citet*{RU08.2} and \citet*{GU10.1} did not find the suppression in power of the total matter at intermediate scales that was found instead by \citet{JI06.1}.

Finally, \citet{VA11.1} performed a similar study based on numerical simulations, adopting $2\times512^3$ gas and dark matter particles within a box of $100~h^{-1}$ Mpc on a side, and introducing (for the first time) the effect of energy feedback from Active Galactic Nuclei (AGN henceforth) to the physics implemented by earlier simulations. First of all, they found a percent level decrease in the clustering amplitude of total matter at intermediate scales, and a subsequent increase at small scales, like \citet{JI06.1} and unlike \citet*{RU08.2} and \citet*{GU10.1}. They also found that the decrement is much enhanced by AGN feedback, since more gas is pushed to large scales, and since star formation is suppressed, the increase at small scales is also suppressed. Finally, similarly to \citet{JI06.1}, they studied the backreaction of gas onto dark matter finding, in the absence of AGN feedback, a monotonic increase in the dark matter clustering amplitude, up to $\sim 10\%$ at $k\sim 10~h$ Mpc$^{-1}$.

A study of the impact of baryonic physics on the matter power spectrum was also presented in \citet{CA11.1} ($2\times 256^3$ gas and dark matter particles within a $256~h^{-1}$ Mpc side box, modeling radiative cooling, star formation, supernova feedback and UV background). They did not find any reduction in the total matter correlation strength at intermediate scales, contrary to \citet{JI06.1}. Moreover, the increase at small scales of the total matter power spectrum was $\sim 40\%$ at $k\sim 10~h$ Mpc$^{-1}$, intermediate between the results of \citet{JI06.1} and those of \citet*{RU08.2,GU10.1}. 

\subsection{Dynamical DE}

One of the most recent studies concerning how a quintessence-dominated background affects the large scale matter distribution in the Universe is the already mentioned \citet{CA11.1}. There the authors studied the matter power spectra in dynamical DE cosmologies, employing previous results \citep*{LI05.2,FR07.1,CA09.1}. Namely, the non-linear matter power spectrum in a cosmological model with dynamical evolution of DE can be obtained from the non-linear matter power spectrum of a model with a constant $w$ (although depending on the redshift of interest). This phenomenological recipe gives results that are accurate at better than $\sim 1\%$, for the dynamical DE models where it has been tested. The authors found that in a cosmological model where the DE equation of state parameter evolves according to the simple recipe of \citet{CH01.1}, the gas power spectrum is increased at all scales, especially at very small scales, $k\sim 10~h$ Mpc$^{-1}$. The dark matter only spectrum is also increased at all scales, although to a lesser extent, and the total matter power spectrum is slightly increased at intermediate scales ($\sim$ a few $h$ Mpc$^{-1}$) and slightly decreased at small scales. Unfortunately, the authors do not go to a great length in discussing the physical interpretation of their results, being mainly interested in showing that their phenomenological recipe works also with baryonic matter components.

Earlier efforts performed in order to understand the impact of a time varying equation of state parameter for DE (we exclude the studies dealing only with a constant $w$, or where the quintessence field is coupled to matter) on the large scale matter distribution never included the effects of baryons, and date back to \citet{KL03.1}. They performed simulations with varied box sizes and mass resolutions, considering the \citet{RA88.1} standard quintessence model (see Section \ref{sct:models} below) with a $\sigma_8$ matching the value for the fiducial $\Lambda$CDM cosmology, finding a substantial increase in power at scales $k\gtrsim 5~h$ Mpc$^{-1}$, especially at high redshift.

More recently, \citet{MA07.3} performed additional $n-$body simulations using several different recipes for $w(z)$, including the simple parametrization of \citet{CH01.1}. She finds qualitatively different results from \citet{KL03.1}, in that the ratio of the dark matter power spectrum to its counterpart in the fiducial $\Lambda$CDM cosmology has a depression at mildly non-linear scales $k\sim 0.5-1~h$ Mpc$^{-1}$ and a subsequent increase at small scales. The difference with respect to the previous work might possibly reside in the fact that she lets the power spectrum normalization $\sigma_8$ be lower in the various DE models, in order to compensate for the enhanced growth factor. As a matter of fact, when the author imposes the linear matter power spectrum at $z=0$ to be the same as in the fiducial $\Lambda$CDM cosmology, her results are in more agreement with those of \citet{KL03.1}.

Finally, in the aforementioned \citet*{FR07.1} the authors considered again the same \citet{CH01.1} parametrization of the DE equation of state parameter (their simulations have a $256~h^{-1}$ Mpc box side length with $256^3$ particles) and, matching the $\sigma_8$ and the angular diameter distance to the last-scattering surface, found a sub-percent level deviation from the fiducial $\Lambda$CDM case at intermediate scales $k\sim 3~h$ Mpc$^{-1}$ and $z=0$. These results are dissimilar from \citet{MA07.3}, and more in accordance, at least qualitatively, with \citet{KL03.1}. The quantitative differences with respect to the latter can be ascribed to the different $w(z)$ that have been used.

In summary, the effect of baryonic physics on the large scale matter distribution has been studied in several works, employing both numerical simulations and semi-analytic modeling. The results broadly agree at the qualitative level, however they display substantial quantitative differences. Our simulations, described in Section \ref{sct:simulations}, have a larger box size than many of the works detailed above and are hence less prone to finite volume effects. For a few of these works we also employ a better mass resolution, and can thus better resolve small scale physics such as the star formation process. This can be useful in solving, e.g., the ambiguity about the slight drop at intermediate scales in the total matter power spectrum. The impact of dynamical DE on structure formation is much less established. First of all, many studies used the very simple parametrization for $w(z)$ of \citet{CH01.1}, which is not physically satisfactory since it produces an infinite DE density at high redshift unless the DE itself crosses the phantom barrier. Second, the results depend substantially on how the other cosmological parameters are set. For instance dynamical DE models with the same $\sigma_8$ as the fiducial $\Lambda$CDM cosmology are likely to spoil the amplitude of the CMB fluctuations \citep*{FR07.1}. In the present work we used a set of physically motivated quintessence cosmologies with substantially different growth histories \citep{DE10.2}. In all cases the $\sigma_8$ value was adjusted so that the CMB fluctuations amplitude is conserved at the value measured by WMAP. Finally, and most importantly, in this paper we studied precisely the co-evolution of dark matter and baryons in dynamical DE cosmologies, a goal that has never been accomplished at this level of detail and with such an accurate modeling of the gas physics.

\section{Cosmological models}\label{sct:models}

The simulations we investigated in this work span a wide range of flat DE cosmologies. As a reference fiducial model we adopted the standard $\Lambda$CDM model, with parameters in agreement with the WMAP$-3$ years data release \citep{SP07.1}. Accordingly, the matter density parameter is $\Omega_{\mathrm{m},0} = 0.268$, the dark energy density parameter is $\Omega_{\Lambda,0} = 1-\Omega_{\mathrm{m},0}$, and the baryon density parameter is $\Omega_{\mathrm{b},0} = 0.044$. The Hubble constant reads $H_0 = h 100$ km s$^{-1}$ Mpc$^{-1}$, with $h=0.704$. Finally, the primordial power spectrum normalization is set by $\sigma_8 = 0.776$, while its slope is fixed to $n_\mathrm{S} = 0.947$. In what follows, quantities related to this fiducial model will always be labeled by the suffix F. The main difference between the cosmological interpretation of the WMAP$-3$ year data release and those of subsequent releases is in the value of the power spectrum normalization, which tends to be relatively larger for the latter as compared to the former. However we note that the value of $\sigma_8$ adopted here is compatible at $68\%$ Confidence Level with the value extracted from the latest WMAP$-7$ release \citep{KO11.1}.

We then considered four different models involving a dynamical evolution of the DE component. For all these models the cosmological parameters are the same as in the fiducial scenario, except for $\sigma_8$ that was rescaled in a way such that the linear matter power spectrum is always CMB-normalized. Specifically, if $D_+(z)$ is the linear growth factor of a specific cosmology, and $g(z) \equiv (1+z)D_+(z)$, then

\begin{equation}
\sigma_8 = \sigma_{8,\mathrm{F}}\frac{g_\mathrm{F}(z_\mathrm{CMB})}{g(z_\mathrm{CMB})},
\end{equation}
where $z_\mathrm{CMB} = 1089$. It is important to note that, since the growth factor in quintessence cosmologies is always larger than its fiducial counterpart (see \citealt{DE10.2}), the power spectrum normalization will always be such that $\sigma_8 < \sigma_{8,\mathrm{F}}$.

We now proceed to present in more detail the four dynamical DE cosmologies that we investigated. The first two are standard quintessence models, in which the DE is just the energy density associated with a minimally coupled scalar field rolling down some potential \citep*{RA88.1,WE88.1}. The other two instead arise from a non-minimal coupling of the scalar field with gravity, giving effectively origin to the so-called scalar-tensor gravity theories. In what follows we briefly describe both pairs of models separately.

\subsection{Standard quintessence}

In standard quintessence models the evolution of the scalar field $\varphi$ whose energy density represents DE is governed by the Klein-Gordon equation (we work in natural units, $c = \hbar = 1$ unless otherwise noted)

\begin{equation}\label{eqn:klein}
\ddot{\varphi} + 3H\dot{\varphi} + \frac{dV}{d\varphi} = 0,
\end{equation}
which is a direct consequence of the Friedmann equations. In Eq. (\ref{eqn:klein}) a dot denotes derivative with respect to cosmic time, while $V(\varphi)$ is the potential of the scalar field.

As can be seen,  the evolution of the scalar field $\varphi$, is determined by the expansion rate of the Universe, which in turn depends on its whole matter-energy content, and by the effective potential of the scalar field. The two standard quintessence models considered in this work arise from two different choices for the dependence of $V$ on $\varphi$. In the first case a power-law dependence is adopted, according to which

\begin{equation}
V(\varphi) = \lambda^{4+\alpha}\frac{1}{\varphi^\alpha}
\end{equation}
\citep{RA88.1}, where both $\lambda$ and $\alpha \ge 0$ are free parameters. This model will be labeled as RP in the remainder of this work. The second model arises from a generalization of the RP potential, suggested by supergravity considerations \citep*{BR00.1}. In this case the potential reads

\begin{equation}
V(\varphi) = \lambda^{4+\alpha}\frac{1}{\varphi^\alpha}\exp\left(4\pi G \varphi^2\right),
\end{equation}
and will be labeled as SUGRA henceforth.

\subsection{Scalar-tensor theories}

For the second class of models the scalar field $\varphi$ is coupled non-minimally with gravity, giving rise to an effective scalar-tensor gravity theory. We call these models \emph{extended quintessence} models. In order to be more specific, a new function $f(\varphi,R)$ is to be introduced in the gravity action, which describes the coupling between the quintessence scalar field and the Ricci scalar $R$. In the extended quintessence cosmologies considered in this paper, we assume that the coupling function can be factorized as 

\begin{equation}
f(\varphi,R) = R\left[ 1+8\pi G_*\xi\left(\varphi^2 -\varphi_0^2\right)\right].
\end{equation}
In the previous equation, $G_*$ is a 'bare' gravitational constant, that is in general different from Newton constant $G$ and it is set in order to match local constraints on General Relativity.

The strength of the coupling is defined by the parameter $\xi$. Particularly, in this paper we have considered two extended quintessence models, one with $\xi < 0$ (specifically $\xi=-0.072$, EQn henceforth), and one with $\xi>0$ ($\xi=+0.085$, EQp). The potential of the scalar field, although not the only source term of the resulting Klein-Gordon equation, needs to be specified in this case as well. Here, we adopted the RP power-law potential described above. It is worth noticing that, while the 'bare' gravitational constant $G_*$ is a constant of the theory, the effective gravitational constant that is experienced by particles in the simulation is a function of cosmic time. Specifically, it converges to $G_*$ at $z=0$ and approaches a constant value larger (smaller) than $G_*$ at high redshift for positive (negative) $\xi$.

\section{Numerical simulations}\label{sct:simulations}

The numerical simulations are fully described in \citet{DE10.2}. Here we just recall the main features, referring the interested reader to that work for further details. The evolution of the large-scale matter distribution in the various DE cosmologies was simulated by using the {\sc Gadget-3} code (\citealt*{SP01.1}; \citealt{SP05.1}). This code allows to simulate the evolution of both the dark matter and the gas component, making use of the entropy conserving formulation of SPH \citep{SP02.1}. The code was suitably modified in order to allow for an arbitrary time evolution of the DE equation of state (see for instance \citealt{DO04.1}), and also for an arbitrary time evolution of the effective gravitational constant, needed for the extended quintessence models.

The baryonic physics implemented in the simulations includes radiative cooling, heating by an uniform redshift-dependent UV background \citep{HA96.1}, and star formation and feedback. The latter requires sub-grid modeling of the multi-phase Inter Stellar Medium (ISM), with the cold phase providing the reservoir for the star formation, and the hot phase being fueled by supernova heating. The latter provides also evaporation of some cold clouds, thereby leading to a self-regulation of the star formation process.

Each simulation has a comoving box side measuring $L = 300~h^{-1}$ Mpc and follows the evolution of $768^3$ dark matter particles and an equal number of gas particles. The mass of each dark matter particle is $3.7\times 10^9 M_\odot h^{-1}$, while that of each gas particle equals $7.3\times 10^8 M_\odot h^{-1}$. As in \citet{DO04.1}, the initial redshift is not the same for each simulation, rather for each DE cosmology it is rescaled in order to conserve the linear growth factor at the initial redshift in the fiducial $\Lambda$CDM simulation. Accordingly, all simulations start from the same random phases, but with amplitude of the initial fluctuations rescaled to satisfy the CMB constraints. For every cosmological model, a dark matter only control simulation was ran, providing the reference for the effect of gas physics. In the control runs the number of dark matter particles is $768^3$, each one having a mass of $4.4\times 10^9 M_\odot h^{-1}$. In the following, we will often refer to the dark matter power spectrum in a control run as simply dark matter only power spectrum. 

The power spectra of each matter component in each simulation were computed by projecting the matter distributions on a regular grid constituted by $n=960$ grid points on a side, and then using discrete Fourier transforms. The matter density at every grid point was evaluated by the Triangular Shape Cloud (TSC) method (refer to \citealt{HO88.1}), which was then compensated for interpolation effects (see also \citealt{JI05.1}). The Nyquist frequency corresponds therefore to $k_\mathrm{N}  = (n-1)~\pi/L \sim 10 h$ Mpc$^{-1}$, meaning that the matter distribution is not sufficiently sampled (and hence was not considered in our analysis) on scales $k \gtrsim k_\mathrm{N}$. The simulations we used were not re-run adopting mass resolutions different from the nominal one, and hence we have no way to straightforwardly asses the robustness of our power spectra analysis with respect to particle number and/or box size. Nonetheless, e.g., \citet{BO06.2} shown that changes in the mass resolution have a negligible impact on the simulated distribution of baryons at cluster scales, except for a change in the star formation rate at $z \gtrsim 3$, thus hinting that this should not be an issue for the present work.

\section{The halo model}\label{sct:halo}

In the course of our analysis we wanted to have a semi-analytic prescription for computing the dark matter power spectrum, in order to physically interpret our results. While fits to numerically simulated matter power spectra \citep{PE94.1,SM03.1} are widely used, they have two drawbacks for our purposes. First, they do not really have a physical foundation (although the \citealt{SM03.1} prescription uses some elements of the halo model, it is ultimately only a fit), thus they are not useful when trying to interpret our results in a physical framework. Second, these fits have been matched against simulations containing dark matter only and in a $\Lambda$CDM background. Since they lack a physical motivation, there is no telling what is their range of validity for alternative cosmologies (see however \citealt{MA07.3}), or even for scales outside the range probed by those simulations. As a consequence, we adopted the physically motivated halo model.

The halo model \citep{MA00.3,SE00.1,CO02.2} is a physically motivated framework for describing the large scale matter distribution. It is based on the assumption that each matter particle in the Universe is locked inside halos of some mass. If this is true, then the correlation function of matter particles is given by the sum of two contributions. The first one is the contribution of particle pairs belonging to the same halo, and dominates on small scales. The second one is the contribution of particle pairs residing in separated halos, and dominates on large scales. It is natural to expect that, while the $2-$halo term depends on the bias of dark matter halos, the $1-$halo term is more dominated by the internal structure of individual halos. For details on the implementation of the halo model we refer the interested reader to \citet{FE10.1} and references therein. Here we just want to stress the particular importance of the relation between mass and average concentration of cosmic structures. In the present work we adopted an empirical relation that reproduces well the power spectrum fits to numerical simulations \citep{PE94.1,SM03.1} for the $\Lambda$CDM cosmology, that is

\begin{equation}\label{eqn:conc}
c(M,z) = \frac{c_0}{1+z}\left[ \frac{M}{M_*(z)} \right]^\beta,
\end{equation}
where $c_0 = 10$ and $\beta = -0.15$. In the previous equation $M_*(z)$ is the typical collapsing mass at a given redshift, and it is the only cosmology-dependent quantity that enters the concentration-mass relation. This means that when we applied the halo model to cosmologies with dynamical DE, the concentration-mass relation for dark matter halos was modified only through $M_*(z)$. We also observe that the concentration of halos with a given mass decreases faster with redshift according to Eq. (\ref{eqn:conc}) than the $\propto (1+z)^{-1}$ behavior found by, e.g., \citet{BU01.1} or \citet*{EK01.1}. This kind of issue has been already noted and studied in detail by \citet{SE03.2}.

Before proceeding we would like to stress a subtlety of our approach. In the present work we tried to fit ratios between simulated power spectra with ratios between outcomes of the halo model. Although the halo model by itself gives a good description of the true matter power spectrum, it cannot be expected to be precise at the percent level, as are the effects that we are considering here. This is due to the many obvious approximations and simplifications that are introduced in the model. Besides, the prescriptions for the mass function of dark matter halos and their linear bias (that are two ingredients of the halo model) can be considered accurate only at the $\sim 10\%$ level at best. However, by considering only ratios we are presumably canceling out these discrepancies, thus highlighting the true physical effects. 

\section{Results}\label{sct:results}

We now proceed to present our results. We treat at first the effect of baryonic physics in the fiducial $\Lambda$CDM cosmology, in order to compare our results with those of previous works in the field. Then we move on to consider the effect of a dynamical DE evolution on each matter component of the simulations, including both dark matter and baryons. Subsequently we explore the impact of baryonic physics on the dark matter distribution in each quintessence cosmology, and we conclude by exploring the behavior of the total matter power spectrum in these models.

In what follows, we shall always refer to the dimensionless power $\Delta^2(k,z)$, defined as

\begin{equation}
\Delta^2(k,z) = \frac{k^3P(k,z)}{2\pi^2}.
\end{equation}
The dimensionless power is basically the contribution to the mass variance given by logarithmic intervals in wavenumber. As mentioned, in many circumstances we have shown ratios of powers, thus rendering the distinction between the dimensionless power and the power spectrum irrelevant. 

\subsection{Impact of gas physics in the $\Lambda$CDM cosmology}

\begin{figure}
	\includegraphics[width=\hsize]{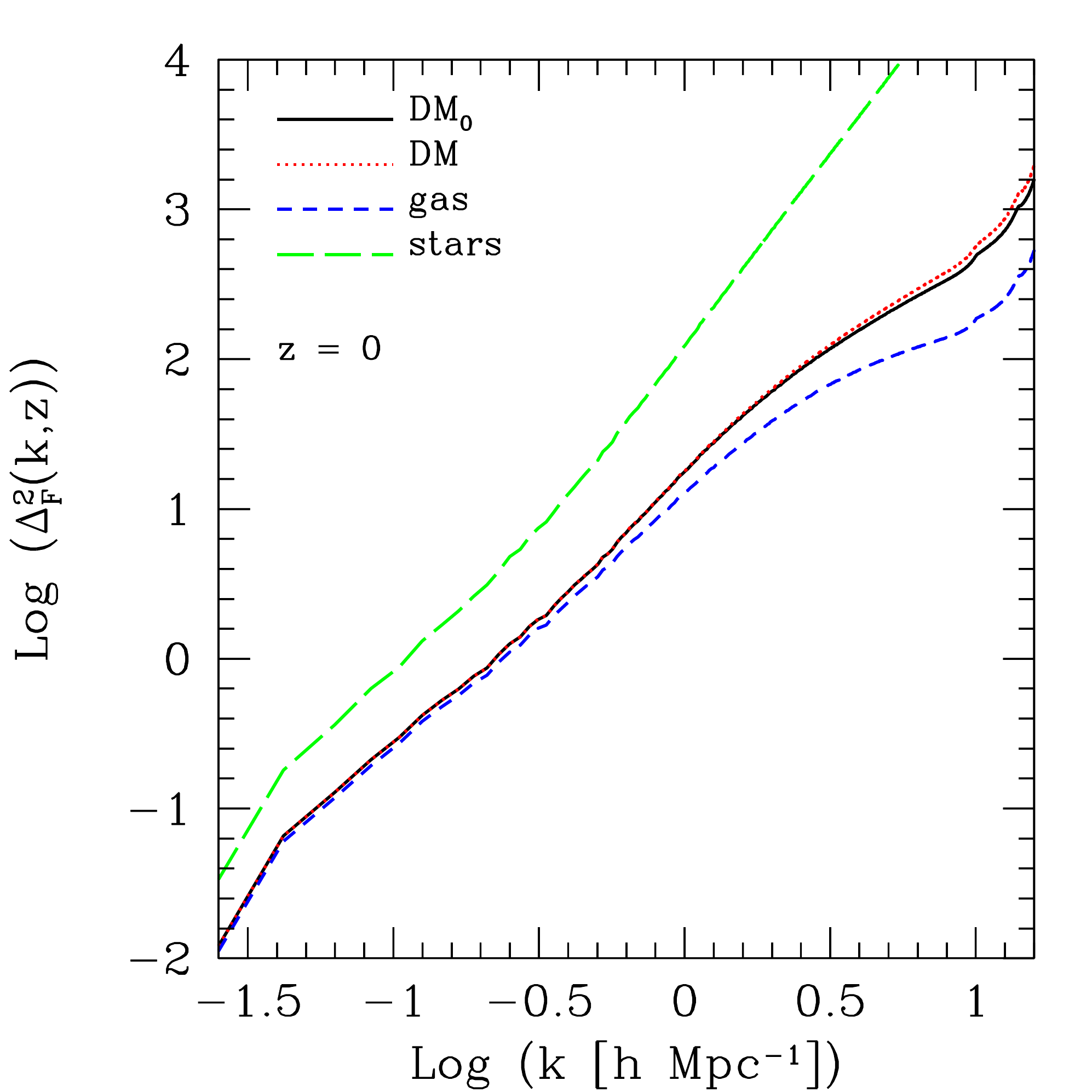}\hfill
	\caption{The dimensionless powers of dark matter, gas, and stars in the fiducial $\Lambda$CDM cosmology. The black solid line refers to the dark matter-only simulation, while the other line types refer to the various matter components in the full simulation. We did not plot the total dimensionless power since on these scales it basically overlaps with the dark matter one. All curves refer to $z=0$.}
\label{fig:powerSpectrum}
\end{figure}

We first turn attention to the power spectrum of the various matter components in the $\Lambda$CDM simulation. For definiteness, in Figure \ref{fig:powerSpectrum} we show the $z=0$ dark matter only power for the fiducial $\Lambda$CDM cosmology control run, as well as dimensionless powers for the different matter components in the full simulation.

The most evident fact about this Figure is the large discrepancy between the power spectrum of the stars and that of the other matter components. This difference is expected, since stars tend to form inside dark matter halos, that are biased tracers of the underlying smooth density field. Therefore stars are more clustered together compared with dark matter or gas particles. A similar behavior for the correlation function of stars in numerical simulations has been found, e.g., in \citet{JI06.1} (see Section \ref{sct:previous} above).

Next, we consider the behavior of gas. As can be seen by looking at Figure \ref{fig:powerSpectrum}, the gas traces quite well the dark matter distribution at large scales, with only very little (and smaller than unity) bias. However, as the scales become more and more non-linear, the clustering strength of gas particles becomes smaller and smaller as compared to that of dark matter. This behavior is due to a combination of star formation and hot gas pressure. Specifically, part of the gas tends to cool down and transform into stars in the cores of dark matter halos, implying a depletion at relatively small scales. At the same time, shock heating and gas pressure prevent the hot gas component to have small scale clustering. Again, similar qualitative conclusions have been reached previously by \citet{JI06.1}, \citet*{RU08.2}, and \citet{VA11.1}.

\begin{figure}
	\includegraphics[width=\hsize]{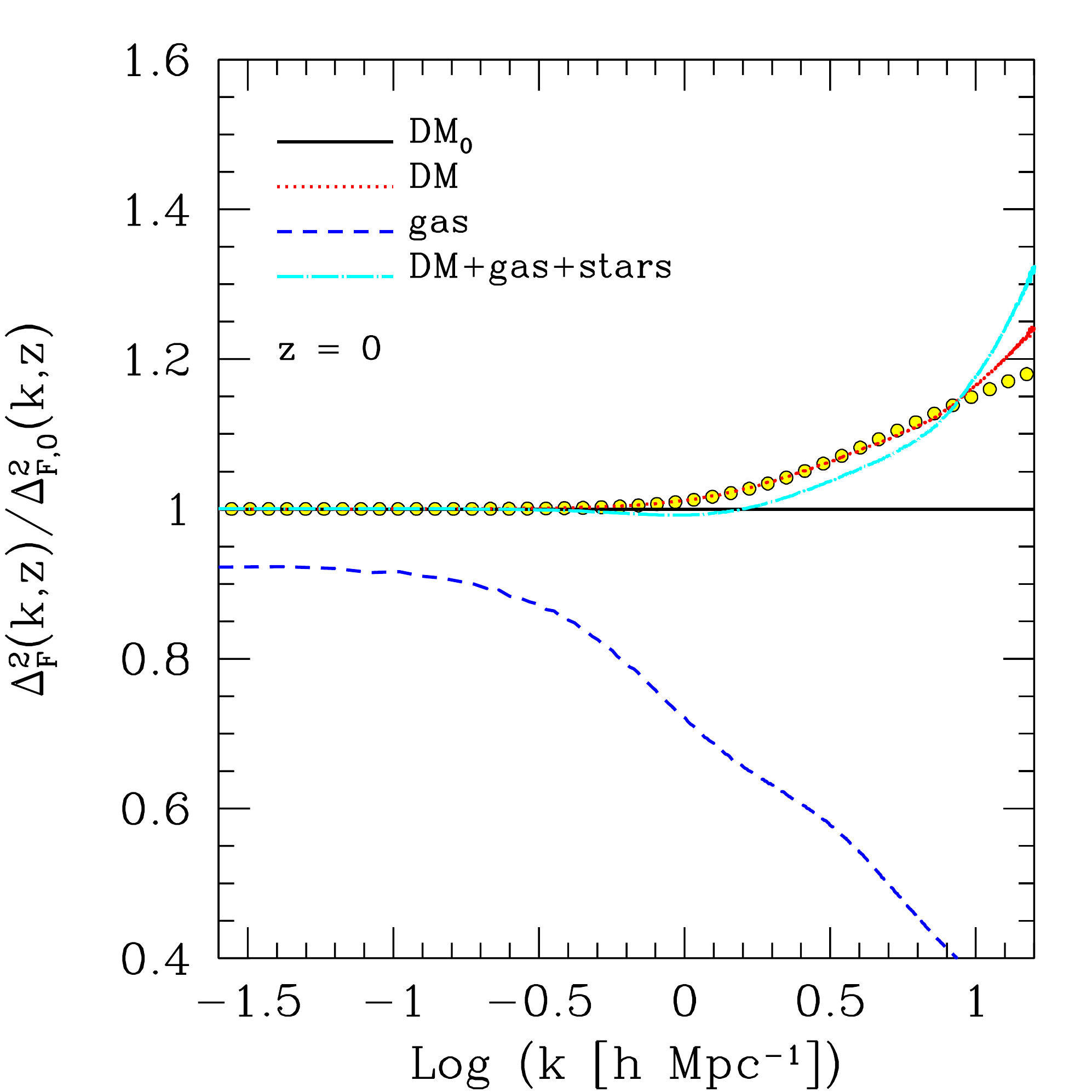}\hfill
	\caption{The ratio between the power spectra of the various matter components in the full simulation (as labeled in the plot) for the fiducial $\Lambda$CDM cosmology and the matter power spectrum for the dark matter only control run. The yellow circles represent the halo model predicted dark matter power spectrum obtained by increasing the average structure concentration according to the recipe detailed in the text, normalized by the original halo model spectrum. All quantities refer to $z=0$.}
\label{fig:powerSpectrumRatio}
\end{figure}

Figure \ref{fig:powerSpectrum} also allows us to understand what happens to the dark matter component upon inclusion of the gas physics, that is the backreaction of gas upon dark matter, which is a central focus of this work. As can be seen, the dark matter indeed gains a little bit of clustering strength at small scales. This can be interpreted in the following way: while the gas cools and concentrates in the very center of dark matter halos, the dark matter particles are also pulled closer to the halo centers. Moreover, adiabatic contraction due to gas radiative losses also contributes to make dark matter halos more concentrated. Overall, dark matter particles inside individual structures tend hence to be more strongly clustered. The same qualitative effect has been found by \citet{JI06.1} and \citet{VA11.1}. We can further investigate this important issue by plotting the ratio of the various spectra to the dark matter only spectrum. This is done in Figure \ref{fig:powerSpectrumRatio}.

In this Figure we did not show the stellar power spectrum, since it would be way out of scale. On the other hand, we added the total matter power spectrum, that in Figure \ref{fig:powerSpectrum} would be too similar to the dark matter one to be visible. In this plot we can better appreciate the different behavior of the gas component as compared to the dark matter one. Particularly, the clustering amplitude of the former is suppressed by up to a factor of $\sim 2.5$ in comparison to the latter at scales $k \lesssim 10~h$ Mpc$^{-1}$. In several previous works (\citealt*{RU08.2}; \citealt{CA11.1,VA11.1}) it can be seen that the gas power spectrum has the tendency to catch up on the dark matter-only spectrum at scales $k \sim 10~h$ Mpc$^{-1}$ or smaller. We do not see trace of this behavior in our plot, although it is possible to happen at scales below the smallest ones probed by our runs. We attribute this discrepancy to the differences in the implementation of gas physics. Finally, as noted by \citet*{RU08.2}, the large scale smaller than unity bias of the gas component with respect to the dark matter component is likely due to the fact that low mass dark matter halos (that are less biased with respect to the large scale matter distribution) contain a larger gas fraction as compared to the high mass (and more clustered) structures.

Let us now focus on the modification to the dark matter power spectrum due to the inclusion of gas physics, that is the backreaction. As can be seen from Figure \ref{fig:powerSpectrumRatio}, the dark matter power is correspondingly enhanced by up to a factor of $\sim 15-20\%$ at the smallest scales probed by the simulations. This is  somewhat larger than what is found by \citet{JI06.1} and \citet{VA11.1} ($\sim 10\%$), and at the same time smaller than the results of \citet{CA11.1} ($\sim 40-50\%$). Again, these differences can be explained in the different details of gas physics implementation that, as shown by \citet{VA11.1} can have a large impact. In Figure \ref{fig:powerSpectrumRatio} we also plot the dark matter power spectrum evaluated through the halo model by assuming a higher average concentration of dark matter halos, normalized by the regular halo model output. More specifically, we modified the concentration of a halo with given mass and redshift according to $c_\mathrm{F}(M,z) = fc_{\mathrm{F},0}(M,z)$, where $f = 1.17$. As can be seen, this very simple recipe accounts very well for the effect observed in the simulations, and the deviations between the halo model prediction and the simulation results are $\lesssim 1\%$ at all scales probed by the latter ($\lesssim 10~h$ Mpc$^{-1}$).

Realistically, one might expect the effect of baryonic physics to act differently on dark matter halos of different masses, with less massive halos being more affected as compared to more massive structures. Nevertheless, simply changing the concentration of all halos of the same amount irrespective of the mass is much simpler and, as can be seen, works remarkably well. The modification to the halo model described above matches also fairly well the total matter power spectrum, represented by the cyan line in Figure \ref{fig:powerSpectrumRatio}, with maximum deviations being of the order of $3-4\%$. As a matter of fact in \citet*{RU08.2}, the authors used halo model considerations and increased the concentration of structures in order to reproduce the total matter power spectrum. In \citet*{GU10.1} the authors did the same by also including a model for the distribution of stars in the very center of dark matter halos. Both works managed to reproduce the ratio of the total matter power spectrum to the dark matter only power spectrum with an accuracy of $\sim 5 \%$ or worse, and deemed this result as acceptable. 

\begin{figure*}
	\includegraphics[width=0.8\hsize]{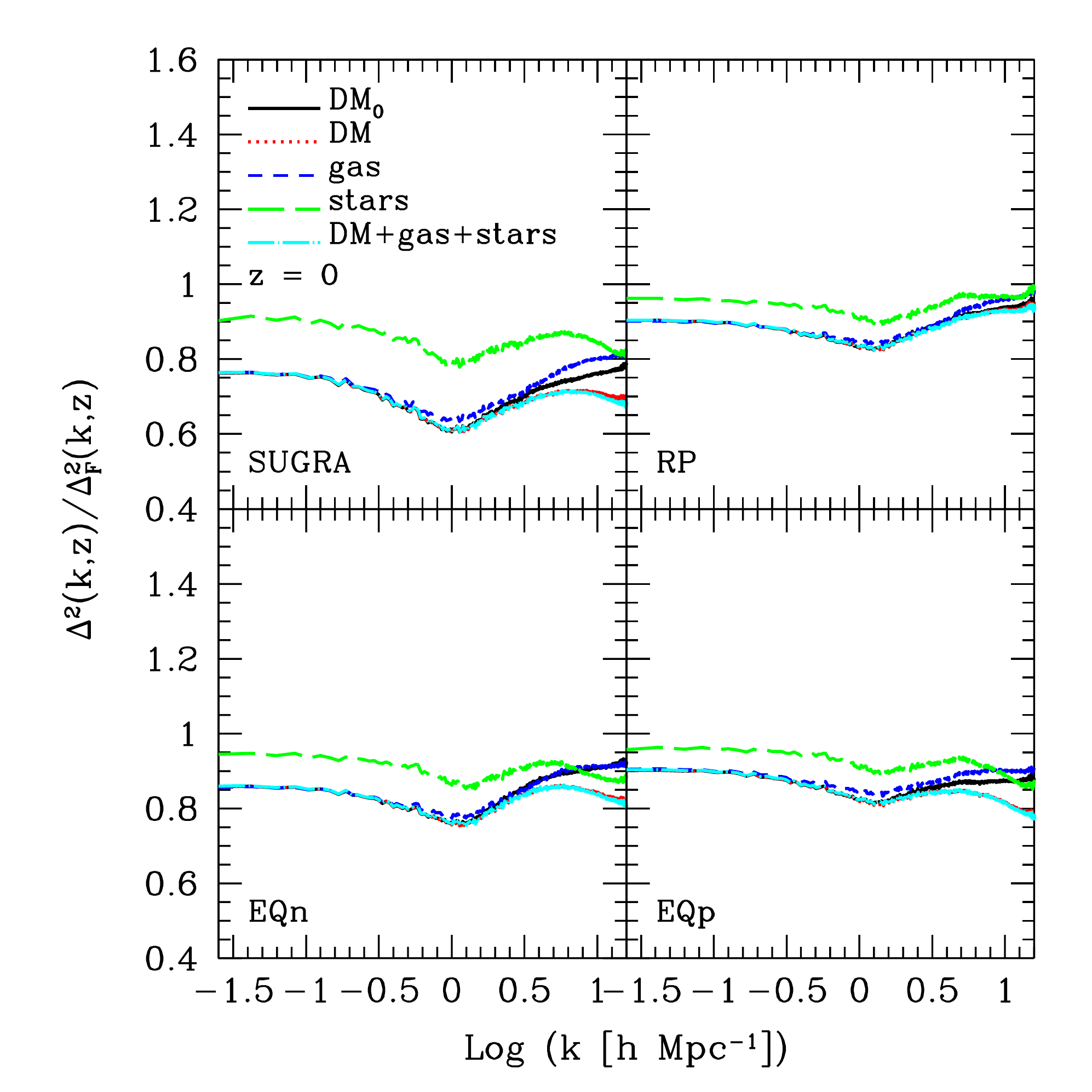}\hfill
	\caption{The ratio between the power spectra of the various matter components in the four quintessence simulations adopted in this work and the corresponding quantities in the fiducial $\Lambda$CDM run. The black solid line refers to the power spectrum of the dark matter only simulation, while the other line types refer to the various matter components of the full simulations, as labeled in the plot. All quantities are evaluated at $z=0$.}
\label{fig:powerSpectrumRatio_DE}
\end{figure*}

It would be interesting if the halo model could be straightforwardly modified in order to reproduce the total matter power spectrum at the sub-percent level of accuracy since, for instance, cosmic shear surveys measure the total power, rather than the dark matter power. However the behavior of the former is more complicated than that of the latter. First of all, there is a slight suppression of power, at the percent level, at intermediate scales. As mentioned in Section \ref{sct:previous} this was not found by \citet*{RU08.2} and \citet*{GU10.1}, but was found by \citet{JI06.1} and, more recently, by \citet{VA11.1}. Even if this is ignored however, the total matter power spectrum tends anyway to be shallower than the dark matter one at intermediate scales and steeper at small scales. This is due to the fact that the gas contribution pushes it down at intermediate scales and the stellar contribution pulls it up at small scales. As a consequence, the behavior of the total matter power spectrum cannot be easily reproduced with a simple modification of the concentration of structures, at least if we want sub-percent level precision. From the halo model point of view, this can be interpreted as the consequence of a twofold effect: $(i)$ the shape of individual structures (containing dark matter, gas, and stars) is not as well described by a NFW profile as are dark matter halos alone; $(ii)$ the collisional nature of gas makes sure that the assumption according to which all matter is locked into structures is not valid. In other words, the halo model should include an additional component describing the fraction of gas that is still smoothly distributed, similarly to what happens for Warm Dark Matter \citep{SM11.1} scenarios. An interesting direction for future investigation would be to develop a generalized halo model that would include the distribution of dark matter, hot and cold gas, and stars. The parameters of such a model could definitely be adjusted to reproduce the total matter power spectrum at the sub-percent level, however it is well beyond the scope of the present paper.

\subsection{Matter power spectra in quintessence cosmologies}

In Figure \ref{fig:powerSpectrumRatio_DE} we show the power spectra of the different matter components in the four dynamical DE simulations considered in this work, normalized by the corresponding quantities in the fiducial $\Lambda$CDM run. Due to the fact that the normalizations of the $z=0$ power spectra in the quintessence models are always smaller than the corresponding quantity in the fiducial model, the clustering strength of all matter components at very large scales is always reduced. The first interesting thing to note is that all matter components have the same generic shape, being flat at large scales, as expected, with a depression at $k \sim 1~h$ Mpc$^{-1}$ and then growing again at smaller scales. This aspect is reminescent of the dark matter power spectrum shape in dynamical DE models explored by \citet{MA07.3}, with the difference that there the depression was located at slightly larger scales. As a matter of fact, even here we note that the dip is at slightly smaller scales in the models RP and EQp as compared to the models SUGRA and EQn, suggesting that while the presence of the depression is a generic feature, its precise location depends on the exact background expansion history of the Universe and on the amplitude of primordial density fluctuations. This is supported by the fact that, as we will discuss shortly with the help of the halo model, the scale of the dip depends on the scale at which non-linear effects become important. It should also be noted that when the value of $\sigma_8$ is adjusted to be the same for all models the depression is expected to disappear, as happen in the results of \citet*{FR07.1} and \citet{CA11.1}.

Let us next examine the behavior of the gas component. The relevant power spectrum ratio is equal to the ratio of the dark matter power spectra (both in the dark matter only and in the full simulations) at large scales. However, at intermediate/small scales it tends to be larger. This means that, while the gas clustering amplitude in quintessence cosmologies remains smaller than its $\Lambda$CDM counterpart, $(i)$ less gas is depleted and locked into stars in cosmologies with a dynamical evolution of DE, and $(ii)$ shock heating and hot gas pressure are slightly less efficient in preventing gas clustering. Which one of these two effects is the dominant one is hard to say, however we note that \citet{DE10.2} already pointed out that star formation tends to be suppressed in dynamical DE cosmologies as compared to the fiducial $\Lambda$CDM model. An increase in the gas clustering strength at small scales for dynamical DE models has furthermore been found by \citet{CA11.1}, where the authors also note that star formation is reduced. Curiously, this means that the reduction in the star formation efficiency is not (or not completely) to be attributed to the smaller $\sigma_8$ values adopted here. All in all this fact makes sense, because a dynamical evolution of dark energy generically implies a larger Hubble drag, and thus an earlier suppression of structure formation. It is plausible that this also translates into a suppression of the star formation process.

The above mentioned conclusion is also supported by observing the stellar power spectrum. Specifically, the ratio of the stellar power spectra is always larger than the ratio of the power spectra for the other matter components, in particular dark matter. This means that the clustering strength of the stars is less suppressed in dynamical DE simulations with respect to the clustering strength of dark matter. In other words, in the latter simulations the stars tend to form into more biased halos as compared to the fiducial run. This is consistent with gas cooling and star formation being more difficult, and hence effective only in more massive halos as compared to the fiducial run.

Finally, it is interesting to note that at small scales the ratio between the dark matter (and total matter) power spectra in the full simulations tends to be smaller than the ratio of the dark matter spectra in the dark matter only runs. This hints to the fact that the dark matter power spectrum is less affected by the baryonic physics with respect to the fiducial $\Lambda$CDM run, again confirming the aforementioned conclusions. Summarizing this part, we can conclude that evolving structures in a quintessence cosmology have an overall qualitative effect on the matter distribution that does not depend much on the specific equation of state parameter $w(z)$ adopted, although the details do, especially at small scales. It is also interesting to note that the extended quintessence models, despite having an effective modification of the gravitational constant, do not behave much differently from the others, standard quintessence models.

In order to better understand the depression in the power spectrum ratios of all matter components that is visible at $k \sim 1~h$ Mpc$^{-1}$ for all quintessence models, we tried to model the ratios between the power spectra in the dark matter only simulations with dynamical DE and the corresponding quantities for the fiducial run by using the halo model. Preliminarily, we should recall that the halo model has some intrinsic dependence on cosmology, through the linear matter power spectrum, the halo mass function and the corresponding linear bias. Moreover, the relation $c(M,z)$ also has a cosmology dependence due to $M_*(z)$. All these dependences have been suitably taken into account, by adopting the correct $\sigma_8$ and growth factor for computing the mass variance and by solving the spherical collapse equations with quintessence contributions. However, the generic shape of the \citet{SH02.1} mass function and of the \citet*{SH01.1} linear halo bias were left intact.
 
\begin{figure}
	\includegraphics[width=\hsize]{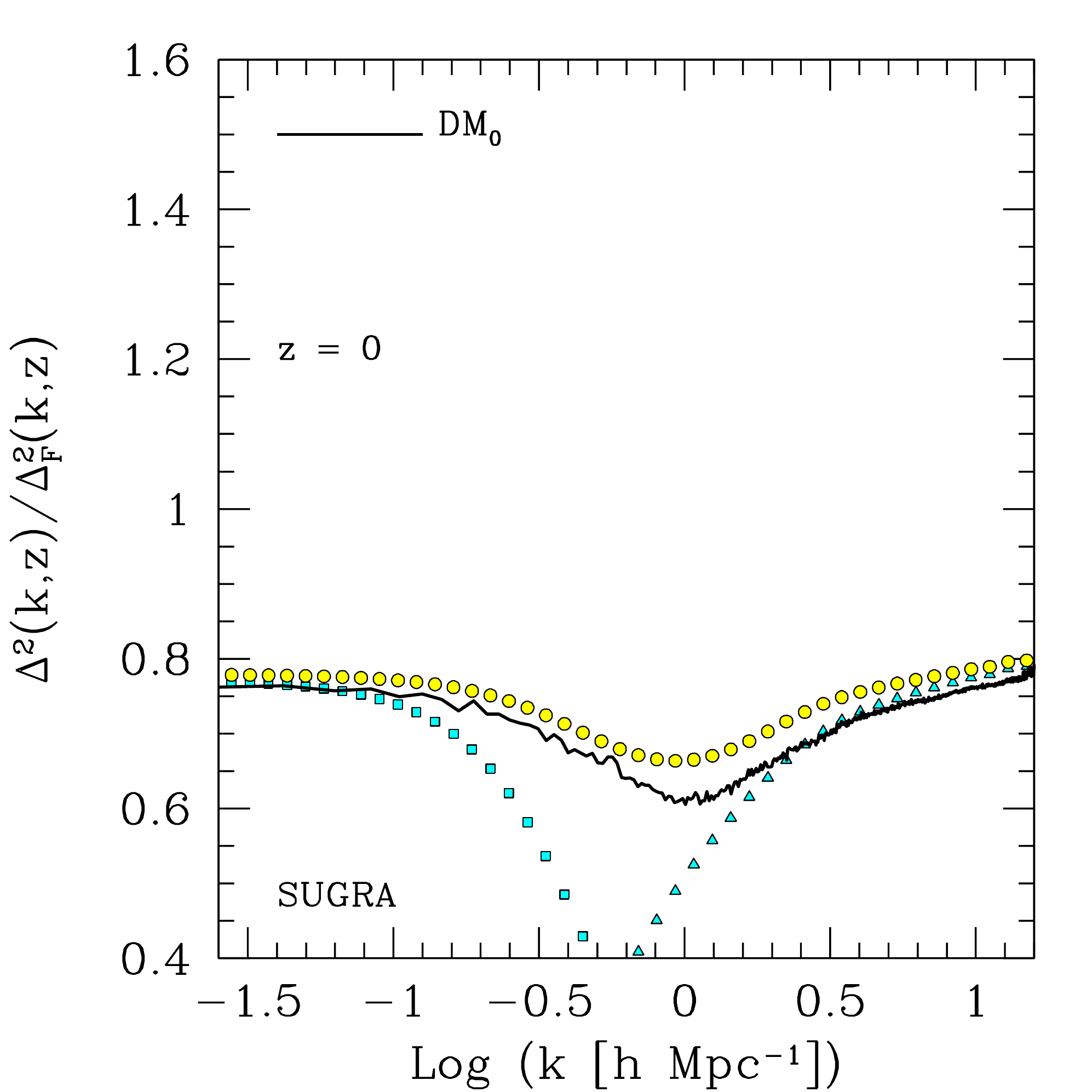}\hfill
	\caption{The ratio between the dark matter power spectrum in the SUGRA quintessence model to the corresponding quantity in the $\Lambda$CDM control run. The black solid line refers to the dark matter only simulation, as labeled. The yellow circles represent the halo model prediction, broken in the $1-$halo (cyan triangles) and $2-$halo (cyan squares) contributions (see the text for more details).}
\label{fig:powerSpectrumRatio_DE_HALOMODEL_UNCORRECTED}
\end{figure} 
 
\begin{figure*}
	\includegraphics[width=0.8\hsize]{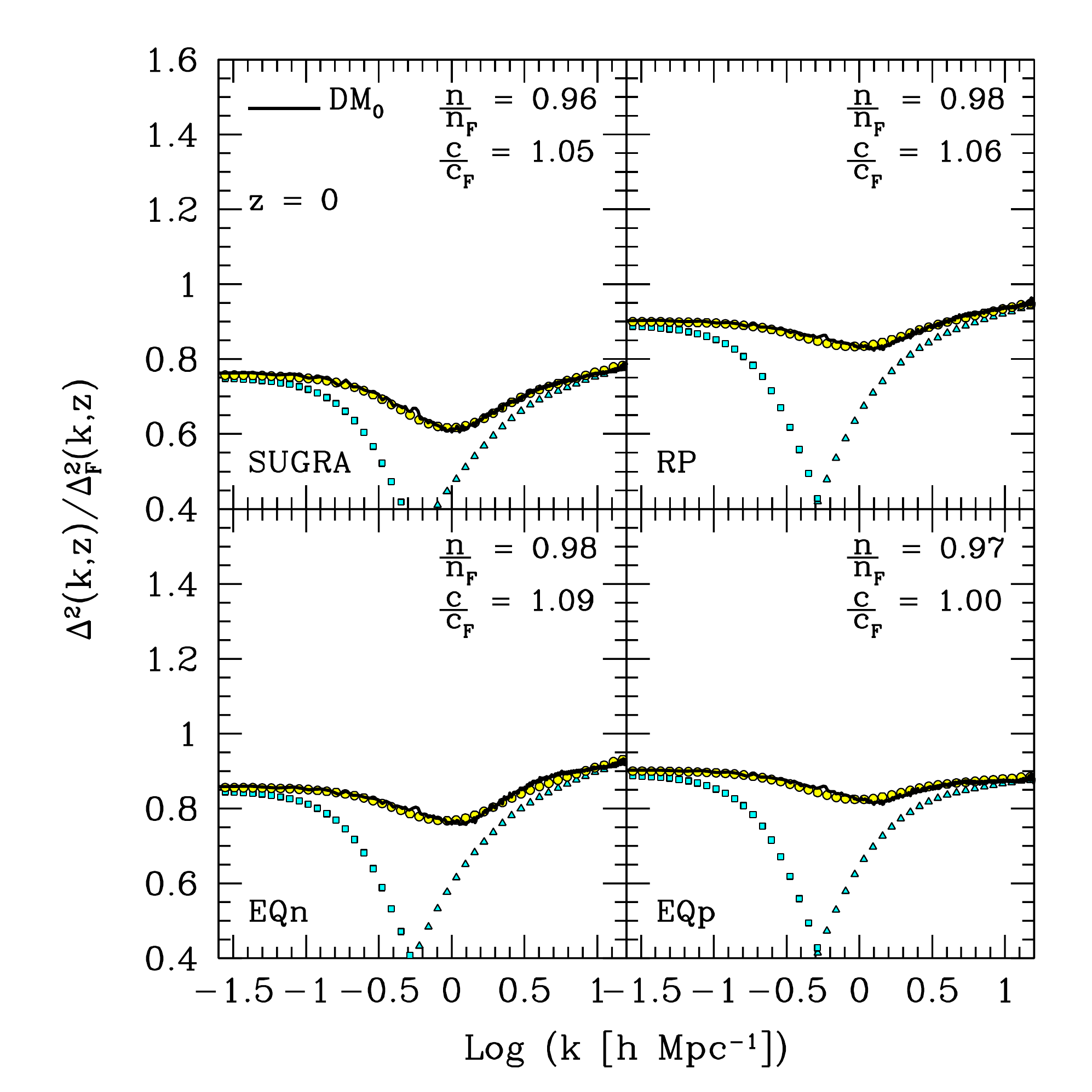}\hfill
	\caption{The ratio of the dark matter power spectra in the four dynamical dark energy cosmologies studied in this work to their counterpart in the fiducial cosmology. The black line refer to the dark matter only run. The yellow circles represent the halo model predictions relative to the black lines, where the three corrections detailed in the text have been taken into account. Particularly, the changes necessary to the mass function ($n/n_\mathrm{F}$) in order to get agreement at intermediate scales and to the average halo concentration ($c/c_\mathrm{F}$) to get agreement at small scales are labeled in each panel of the plot.}
\label{fig:powerSpectrumRatio_DE_HALOMODEL}
\end{figure*}
 
In Figure \ref{fig:powerSpectrumRatio_DE_HALOMODEL_UNCORRECTED} we report the same as in Figure \ref{fig:powerSpectrumRatio_DE} but only for the SUGRA quintessence model and for the dark matter power spectrum in the dark matter only control run. In addition, we overplot the prediction given by the halo model obtained by blindly correcting for cosmology only as described above. We also report the two contributions to the halo model power spectrum, the $1-$halo term and the $2-$halo term, again in units of the dark matter power spectrum in the reference $\Lambda$CDM cosmology. There are a few interesting features to be noted about this plot. First, the overall shape of the power ratio is recovered, and in particular the depression that is visible at $k \sim 1~h$ Mpc$^{-1}$. We can also understand the origin of this depression. As one might expect, the $2-$halo term is a monotonically decreasing function of the wavenumber, while the $1-$halo term is a monotonically increasing function. Therefore, it is natural that the sum of the two displays a minimum, although this minimum does not correspond to the point in which the two terms have equal contributions because they also have a different steepness. 
  
Next, we note that the halo model representation is not entirely accurate from the quantitative point of view, for three reasons: $(i)$ at very large scales the power spectrum ratio evaluated by the halo model is slightly larger than its simulated counterpart, by about $\sim 2\%$; $(ii)$ at intermediate scales this overestimation worsens, or in other words, the dip in the ratio is not accurately reproduced; $(iii)$ at small scales the halo model prediction has a different slope from the simulations, so that even if the former was normalized down, we would still miss agreement. The same conclusions apply to the other three quintessence cosmologies. We now explore each of these issues in order, explaining how we tackled each of them in order to obtain a more accurate description of the dark matter power spectrum in dynamical DE cosmologies.

First of all, we believe the disagreement at large scales not to be of physical, rather of numerical origin. At those large scales the ratio of the power spectra should equal the ratio of their primordial normalizations, that is the ratio of the respective $\sigma_8$ values squared. While this is true for the halo model, the simulated spectra are $\sim 2\%$ off. The only reason for this is that the growth experienced by structures in the simulation is different from the theoretical growth factor. As a matter of fact, we verified that this is the case for all the quintessence models at the $\sim 1\%$ level, which is perfectly acceptable in terms of numerical precision, but that produces the observed offset. In order to compensate for this, we slightly shifted the theoretical $\sigma_8$ values adopted, in order to get a match at large scales.

\begin{figure*}
	\includegraphics[width=0.8\hsize]{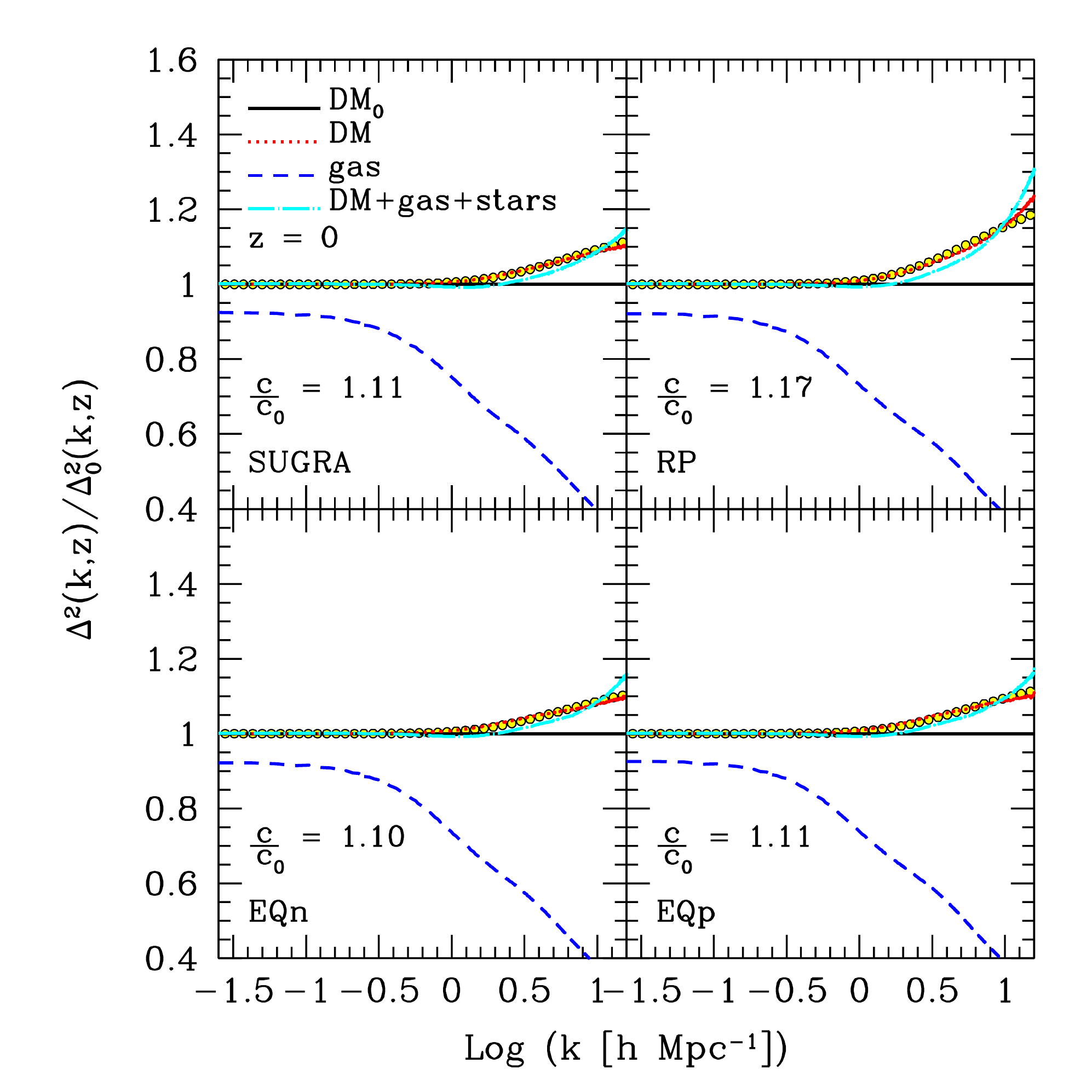}\hfill
	\caption{The ratio between the power spectra of the various matter components in the full simulations (as labeled) and the dark matter power spectra in the dark matter only control runs. Each panel refer to a different quintessence cosmology that has been studied in this work. The yellow circles represent the halo model predictions assuming that the average dark matter halo concentration is increased with respect to the dark matter only case by the quantity labeled in each panel as $c/c_0$.}
\label{fig:powerSpectrumRatio_GAS}
\end{figure*}

Second, the discrepancy at intermediate scales is somewhat expected, since the halo model is well known not to be percent level accurate at the transition between the $1-$halo and the $2-$halo term. It is very well possible that this accuracy actually depends on cosmology, and hence we see a larger discrepancy at those scales. There exist prescriptions that allow to improve the agreement between the halo model and numerical simulations at intermediate scales, for instance the so-called halo exclusion. It means that one should exclude from the computation of the $2-$halo term dark matter particle pairs that are in different but overlapping halos. In other words, one should remove the correlations which arise on scales smaller than the sum of the virial radii of the two halos. At first approximation the halo exclusion manifests in lowering the upper integration boundary in the $2-$halo term \citep{TA03.1,TI05.1}, although more accurate prescriptions exist \citep*{SM07.1,SM11.2}. However, we verified that the halo exclusion does not lead to significant modifications in this case, thus implying that the halo exclusion itself works in quintessence cosmologies as it does in the fiducial $\Lambda$CDM model. On the other hand, we noted that the behavior of the power spectrum ratio at these intermediate scales is quite sensitive to the mass function. By decreasing the mass function for dynamical dark energy cosmologies by only $\sim 3\%$ with respect to their \citet{SH02.1} predictions brings good agreement with the simulated curves. We therefore adopted this minor correction, reminding that even for the fiducial $\Lambda$CDM cosmology the agreement between the \citet{SH02.1} mass function and numerical simulations is only at the $\sim 10\%$ level, and that the level of agreement might very well depend on cosmology. This however has also an interpretation different from the numerical one: the normalization of the \citet{SH02.1} mass function ensures that all the mass in the Universe is locked in to halos (the basic assumption of the halo model). This assumption is of course not strictly true, as some dark matter will certainly be diffused outside halos. Lowering this mass function normalization by the amount discussed means that $\sim 3\%$ less mass in quintessence cosmologies is locked into halos, an interpretation that agrees with structure formation being less efficient in those models.

Finally, the disagreement at the very small scales can be healed by adjusting the dark matter halo concentration, exactly as we did for the effect of baryonic physics. In Figure \ref{fig:powerSpectrumRatio_DE_HALOMODEL} we show the halo model predictions resulting from the three corrections outlined above, for all dynamical DE cosmologies considered in this work. In each panel we also report the corrections to the mass function and to the average concentration of dark matter halos that produce the smallest deviations. With these corrections, the power spectrum ratios given by the halo model reproduce their simulated counterparts with a relative accuracy of $\lesssim 1-2\%$.

As can be seen by the numbers in the various panels, in order to bring the halo model predictions in agreement with the simulations at intermediate scales, the mass function in the quintessence models needs to be lowered by $2-4\%$ at most, a value that is well within the uncertainties of the mass function even for the fiducial $\Lambda$CDM case. After that, in order to match the behavior of the power spectrum ratio at small scales, the average concentration of dark matter halos needs to be increased by $\sim 5-10\%$, with the exception of the model EQp which needs no such correction. In any case, it is likely that the increase in concentration is not entirely physical, rather it is at least in part a response to the mass function decrement, that slightly suppresses the $1-$halo term (not the $2-$halo term, that is normalized at large scales). Since both effects are needed for the overall agreement, it is hard to say what the actual change in concentration would physically be.

\subsection{Gas backreaction in quintessence cosmologies}

In this Subsection we address another tricky point, that is understanding the effect of gas physics on the dark matter power spectrum when the latter is evolved in a background cosmology dominated by dynamical DE. To that purpose, we show in Figure \ref{fig:powerSpectrumRatio_GAS}, the power spectra of the various matter components divided by the dark matter power spectrum in the dark matter only simulation for all the four quintessence cosmologies considered in this work.

The behavior of all matter components looks very similar to that for the reference $\Lambda$CDM cosmology shown in Figure \ref{fig:powerSpectrumRatio}. Upon careful investigation however, it can be noticed that at very small scales the gas tends to be slightly less depleted as compared to the fiducial case, in agreement with the conclusions that we reached before. The most interesting part of this Figure is the comparison of the dark matter power spectrum with the predictions of the halo model. In order to describe the matter power spectrum in dynamical DE cosmologies we adopted the modifications to the mass function and average halo concentrations that were found necessary in the previous Subsection. Then, similarly to what we have done for the $\Lambda$CDM case, we increased the concentrations in order to mimick the effects of baryonic physics.  Since gas cooling and star formation have been proven to be less effective when dark energy is dynamically evolving, we expected to find that an increase in concentration lower than the $17\%$ found for the fiducial cosmology would do the job.

This is indeed our conclusion. With the exception of the RP model, that requires an increase in concentration comparable to the $\Lambda$CDM cosmology, in all other cases and excellent agreement can be reached with an increase of only $\sim 10\%$, as labeled in the various panels of Figure \ref{fig:powerSpectrumRatio_GAS}. In all cases, the relative agreement between the simulated dark matter power ratio and the one predicted by the halo model is always $\lesssim 1\%$. In summary, the halo model with the modifications that we introduced given by the evolution in a quintessence dominated background and the effect of baryonic physics described above reproduces very well gas backreaction on the power spectrum of dark matter, even in cosmologies different from the fiducial one.

The way in which the total matter power spectrum differs from the dark matter only one is rather similar to the fiducial $\Lambda$CDM cosmology. Namely, also for quintessence models we can see a slight suppression of power at mildly non-linear scales, while the behavior at smaller scales cannot be represented by a simple change in the concentration. Nevertheless, if precision at the level of $\sim 3-4\%$ is enough, the enhanced concentration values that we quote in Figure \ref{fig:powerSpectrumRatio_GAS} are a good choice also for the total matter power spectra. In the next subsection we explore with more detail the issue of the depression in the total matter power spectrum at intermediate scales in quintessence cosmologies.

\subsection{The total matter power spectrum}

In this last Subsection we want to focus more on the total matter power spectrum. As we mentioned before, it is not possible to obtain a precise representation of the total matter power by using the halo model in a way as straightforward as for the dark matter power. Nevertheless, understanding the behavior of the total power spectrum is important for weak lensing surveys. The gravitational deflection of light measures the large scale distribution of all matter, irrespective of its nature. Given the large efforts that are being dedicated to future all-sky weak lensing surveys such as \emph{Euclid} \citep{LA09.1} and WFIRST, it is worth studying in more detail this important issue. The total matter power spectra presented in this Subsection, complemented with their redshift evolutions, can be used in order to study the effect of baryonic physics in these four quintessence cosmologies on weak lensing observables.

\begin{figure}
	\includegraphics[width=\hsize]{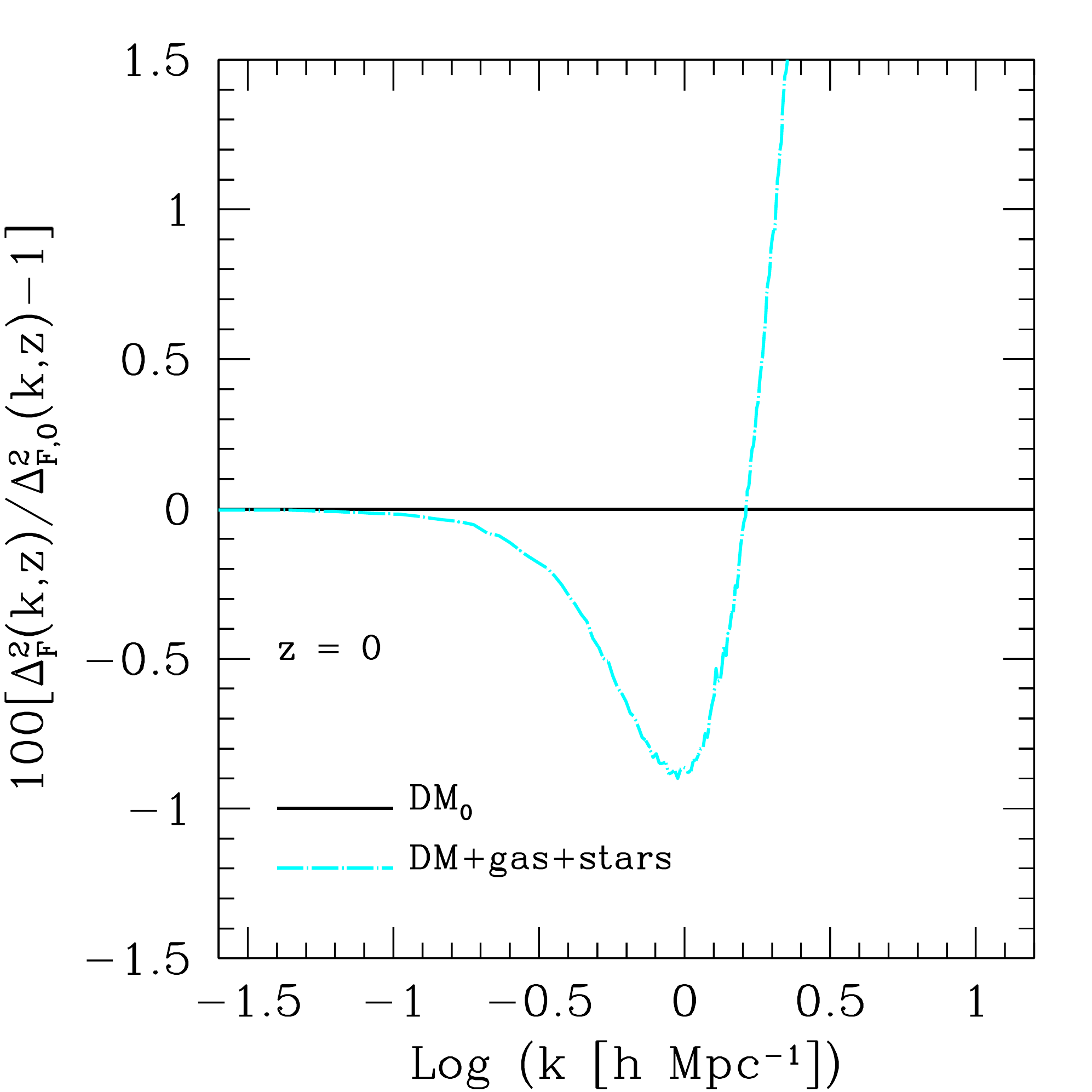}\hfill
	\caption{The total matter power spectrum divided by the dark matter power spectrum in the dark matter only simulation (cyan dot-dashed line). For reference, we also show the dark matter only power spectrum (black solid line). This panel refers to $z=0$ only. Note that the numbers on the ordinate axis express a percentage.}
\label{fig:powerSpectrumRatio_TOTAL}
\end{figure}

\begin{figure*}
	\includegraphics[width=0.8\hsize]{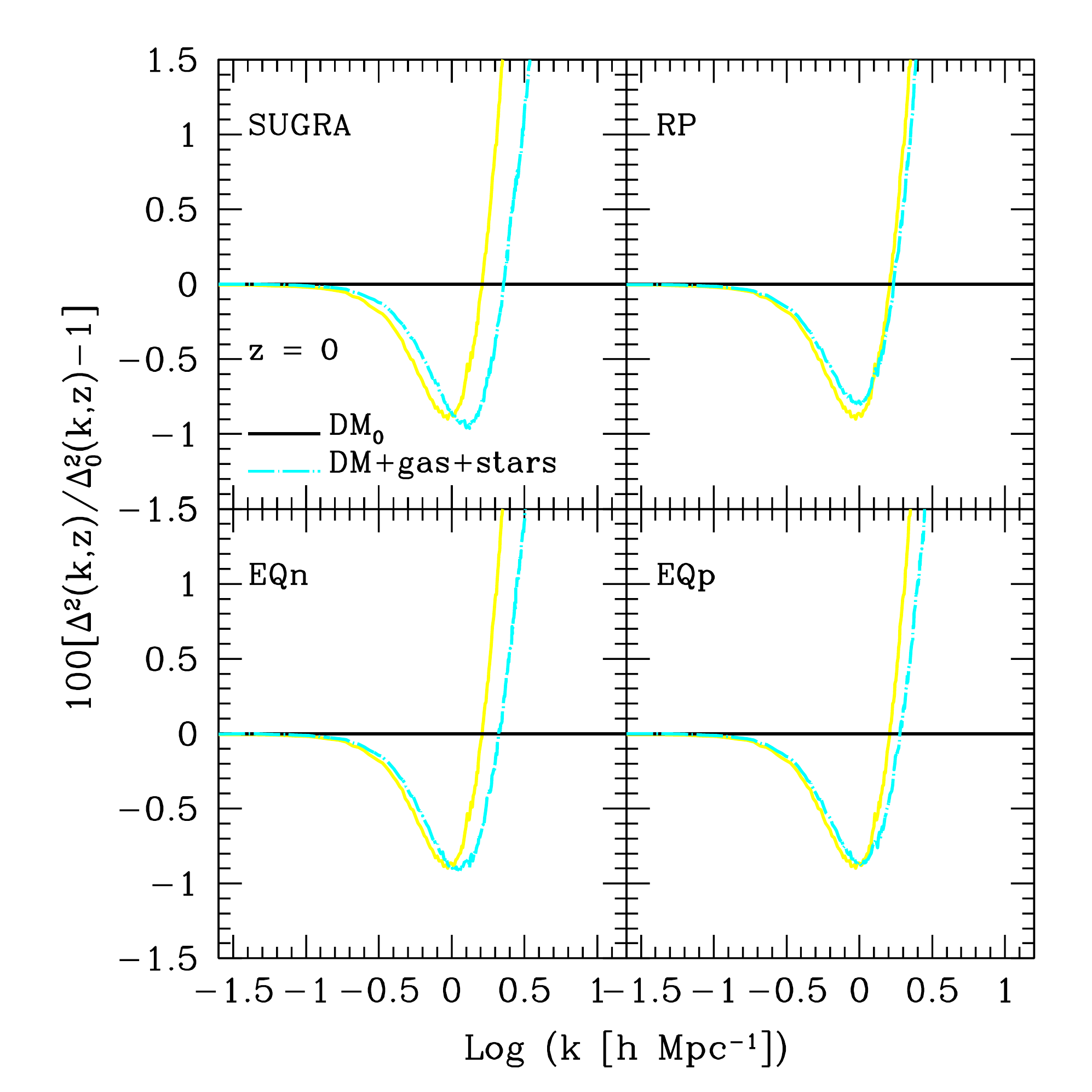}\hfill
	\caption{The total matter power spectra divided by the dark matter power spectra in the dark matter only simulations (cyan dot-dashed lines). For reference, we also show the dark matter only power spectra (black solid lines) and the total power spectrum for the fiducial cosmology, as in Figure \ref{fig:powerSpectrumRatio_TOTAL} (yellow solid line). Each panel refers to $z=0$, and to a different quintessence cosmology, as labeled in the plots. Note that the numbers on the ordinate axis express a percentage.}
\label{fig:powerSpectrumRatio_DE_TOTAL}
\end{figure*}

Let us begin with the fiducial $\Lambda$CDM cosmology. In Figure \ref{fig:powerSpectrumRatio_TOTAL} we show, for this cosmological model, the ratio between the total matter power spectrum and the dark matter power spectrum in the dark matter only simulation. Basically we are showing the same cyan line that is visible in Figure \ref{fig:powerSpectrumRatio}, however this time we report it in percentage, in order to better highlight small differences. As noted above, there is a $\sim 1\%$ depression at scales $k\sim 1~h$ Mpc$^{-1}$. This is in good agreement, both qualitatively and quantitatively, with \citet{JI06.1}. They found that this depression was stable also upon exclusion of the radiative cooling and star formation in the gas simulation, namely by treating the gas as adiabatic. This mean that this feature is not related to the cold gas phase, that condensates and produces stars, rather it should be related with the hot phase. As a matter of fact, the hot phase extends up to intermediate scales, where the effect of adiabatic contraction and of the stellar concentrations are not relevant. There, shock heating and the thermal pressure prevent the gas, and therefore the total matter, from strongly clustering.

The very small amount of the depression in total clustering strength at $k\sim 1~h$ Mpc$^{-1}$ is probably the reason why \citet*{RU08.2} did not detect it in their simulations. The box size that these authors used is relatively small, so that at those scales they were probably still plagued by finite volume effects. For the same reason, simulations using a large box size, like those of \citet{JI06.1} and \citet{VA11.1} did detect this feature, having the same magnitude as here. Despite its very small entity, the shape of this deviation in itself is interesting. If one looks at the effect of primordial non-Gaussianity on the matter power spectrum will find a very similar deviation if the skewness is negative \citep{FE10.1,SM11.1}. The scale of the deviation due to non-Gaussianity is maybe a little bit lower at $z=0$, but most likely has a different redshift dependence than the scale of the one due to baryonic physics. A variation of $\sim 1\%$ as the one induced by gas physics could easily be mistaken with a primordial non-Gaussianity at the level of $f_\mathrm{NL}\sim -50$, if one ignores the information at $k \gtrsim 2-3~h$ Mpc$^{-1}$. Therefore, particular care must be taken when interpreting matter power spectrum results in cosmological terms.

Next, we turn to the quintessence cosmologies. In Figure \ref{fig:powerSpectrumRatio_DE_TOTAL} we show the same as in Figure \ref{fig:powerSpectrumRatio_TOTAL} for each of the four dynamical DE models considered in the present work. The first thing that we note is that the qualitative shape of  the total matter power spectrum is the same as in the reference $\Lambda$CDM cosmology. On the quantitative level, it looks like the depression at intermediate scales is slightly more pronounced and shifted at smaller scales for the SUGRA model as compared to the other quintessence models. This would make sense, because the SUGRA model is also the model where the gas cooling and star formation is more suppressed, hence it is expected to have a larger fraction of gas in the hot phase and stars concentrated at smaller scales. However, the differences are really minuscule, and they overall imply that the distribution and physics of the large scale hot gas are not significantly influenced by the quintessence model. 

\section{Discussion and conclusions}\label{sct:conclusions}

We studied the correlation properties of dark matter, gas, and stars in cosmological models dominated by a dynamical DE component, making use of a suite of large cosmological simulations. The main results of our study can be summarized as follows.

\begin{itemize}
\item In the fiducial $\Lambda$CDM cosmology the gas is much less clustered than the dark matter, more so for small scales, as expected for a part of it to cool down and condense into stars, as well as because of shock heating and gas pressure. The stars are in turn much more clustered than the other matter components, because they form inside biased dark matter halos. The backreaction of gas onto dark matter is such that the clustering strength of the latter is increased by $\sim 20\%$ at small scales ($k\sim 10~h$ Mpc$^{-1}$). The total matter follows a trend very similar (although not identical) to the dark matter.
\item The results of the previous point are in broad overall agreement with previous simulations of structure formation in presence of various gas physical processes not including AGN feedback. Nevertheless, there are a few quantitative differences. Among these, we find a decrement of the total matter power spectrum at intermediate scales $k\sim 1~h$ Mpc$^{-1}$, that some authors find and others do not. The reason for this is probably that the decrement is very small, $\sim 1\%$, at the same order of our own modeling uncertainties. Also, the amount of the increment of clustering strength in the total matter power spectrum at small scales that we find is of $\sim 20-30\%$, intermediate amongst other findings.
\item We used the physically motivated halo model in order to reproduce the backreaction of baryonic physics on the dark matter power spectrum. In particular, we found that by increasing the average concentration of dark matter halos by $17\%$ irrespective of the mass accounts very well for the small-scale increase in dark matter power due to the inclusion of gas and stars. By using this very simple prescription, the ratio of the dark matter power spectrum to its counterpart in the dark matter only simulation is reproduced at better than $1\%$ at all scales of interest.
\item The ratios between various matter power spectra in quintessence cosmologies and their counterparts in the fiducial model have a very specific form that does not depend qualitatively on the precise cosmological model and the matter component. Particularly, these ratios are flat at large scales, have a depression at $k\sim 1~h$ Mpc$^{-1}$, and then raise again at small scales. At least part of this trend is due to the different $\sigma_8$ values adopted in the various models in order to compensate the variations in the growth history.
\item By blindly applying the halo model to the quintessence cosmologies, we manage to reproduce the overall shape of the dark matter power spectrum (normalized by the dark matter only spectrum). However there are a few significant quantitative differences. One of them is that the mass function normalization in these models needs to be artificially lowered by a few percent with respect to the \citet{SH02.1} prediction, in order to match the dark matter power spectrum behavior at intermediate scales. This would mean that slightly less matter is locked into halos in the quintessence models, in agreement with a less efficient structure formation (see below).
\item Quantitatively, the power spectra of the various matter components hint at the star formation process being less efficient in quintessence models as compared to the $\Lambda$CDM cosmology, with the SUGRA model being the least efficient of all. For instance, the gas clustering strength in dynamical DE cosmologies is less decreased, compared to the dark matter one, with respect to the fiducial model, implying that less gas condenses into stars. At the same time, the stars gain more power, implying that they can form only in more massive (and hence more biased) halos. The whole structure formation process seems to be less effective in quintessence cosmologies, as suggested by \citet{DE10.2}.
\item As a further confirmation of this, we also found that the backreaction of the gas onto the dark matter distribution is less severe for the quintessence cosmologies. As it turns out by using the halo model, the average concentration of dark matter halos needs to be increased by $\sim 17\%$ at $z=0$ in the $\Lambda$CDM model (see above), and only by $\sim 10\%$ in quintessence models. The exception to this is the RP model, which is the one with least dissimilarities with respect to the fiducial cosmology.
\item Finally, we also briefly studied the total matter power spectrum, being the subject of cosmic shear observations. When compared to the dark matter only power spectrum, it displays a percent-level reduction of power at intermediate scales, driven by the gas component, and a sharp increase at small scales, driven by the stellar component. Contrary to the dark matter spectrum, the specific behavior of the total matter power cannot be accurately modeled with simple modifications to the halo model.
\end{itemize}

As noted very recently by \citet{VA11.1}, AGN feedback, which was not included in our simulations, might have a substantial impact on the large scale distribution of matter. Particularly, this has the effect of removing large amounts of gas from the centers of clusters and groups of galaxies, thus inhibiting the total matter clustering at intermediate/small scales. The consequence is that the total matter power spectrum would be suppressed more and at larger scales compared to what we have shown in Figure \ref{fig:powerSpectrumRatio_TOTAL}. This also hints to the fact that differences with works where this depression is not noted might depend on the efficiency of the energy feedback from supernovae. The effect of AGN feedback would depend on the co-evolution of supermassive black holes with galaxies, which would likely depend in turn on the expansion history of the Universe. It is therefore not straightforward to understand the effect of AGNs in dynamical DE cosmologies without further simulations. Nevertheless, \citet{VA11.1} show that, for instance, by changing the normalization $\sigma_8$ by a significant amount has a very little effect on AGN feedback.

While this paper was being finalized, a preprint with relevance to the present work appeared on the archive, namely \citet{SE11.1}. There the authors come up with a modification to the halo model similar to the one of \citet*{GU10.1}, specifically overlapping to the dark matter density profile a $\beta-$model for the gas density profile and a point mass representing the central star concentration. They show that in this way they manage to roughly reproduce the total matter power spectrum in units of the dark matter only one, with an accuracy of $\sim 10\%$. Although their work is not yet complete, in that it lacks the backreaction of gas onto dark matter and the smoothly distributed gas component, it is an important step forward in modeling the large scale matter distribution on the Universe.

Despite missing the AGN feedback ingredient, we believe the present work to be a substantial improvement in understanding the large scale gas and star distributions, as well as their backreaction on the dark matter power spectrum. All of these ingredient are going to be of fundamental importance in order to construct a reliable semi-analytic model describing the correlation function of all matter in the Universe. We also provide crucial information about the main differences in structure growth that arise when the background expansion is dominated by quintessence. This is important in order to better understand the differences in the large scale matter distribution that arise due to physically motivated quintessence scenarios, and hence to devise new tests for detecting a possible redshift evolution of the DE equation of state parameter.

\section*{Acknowledgments}

Part of the computations have been performed at the 'Leibniz-Rechenzentrum' with CPU time assigned to the project 'h0073'. LM acknowledges financial contributions from contracts ASI-INAF I/023/05/0, ASI-INAF I/088/06/0, ASI I/016/07/0 'COFIS', ASI 'Euclid-DUNE' I/064/08/0, and PRIN MIUR 'Dark energy and cosmology with large galaxy surveys'. KD acknowledges the support by the DFG Priority Programme 1177 and additional support by the DFG Cluster of Excellence 'Origin and Structure of the Universe'. We are grateful to C. De Boni and V. Pettorino for useful comments on the manuscript, and acknowledge an anonymous referee for insightful observations that improved the presentation of our work.

{\small
\bibliographystyle{aa}
\bibliography{./master}
}

\end{document}